\documentclass[reprint, prd, superscriptaddress, amsmath,amssymb,aps,floatfix]{revtex4-2}
\usepackage[utf8]{inputenc}
\usepackage{hyperref}
\usepackage{url}
\usepackage{xcolor}
\usepackage{braket}
\usepackage[nolist,nohyperlinks]{acronym}
\usepackage{amsthm}

\usepackage{graphicx}
\usepackage{amsfonts}
\usepackage{booktabs}
\usepackage{siunitx}
\usepackage{float}
\usepackage{mathtools}
\usepackage{tabularx}
\usepackage{multirow}
\DeclarePairedDelimiter\floor{\lfloor}{\rfloor}
\usepackage[normalem]{ulem}
\usepackage{orcidlink}
\graphicspath{{figures/}}

\usepackage{natbib}
\begin{document}

\title{Potential impact of noise correlation in next-generation gravitational wave detectors}

\author{Isaac C. F. Wong\,\orcidlink{0000-0003-2166-0027}}
\email{chunfung.wong@kuleuven.be}
\affiliation{KU Leuven, Department of Electrical Engineering (ESAT), STADIUS Center for Dynamical Systems, Signal Processing and Data Analytics, Kasteelpark Arenberg 10, 3001 Leuven, Belgium}
\affiliation{Leuven Gravity Institute, KU Leuven, Celestijnenlaan 200D box 2415, 3001 Leuven, Belgium}
\author{Peter T. H. Pang\,\orcidlink{0000-0001-7041-3239}}
\affiliation{Institute for Gravitational and Subatomic Physics (GRASP), Utrecht University, Princetonplein 1, 3584 CC Utrecht, The Netherlands}
\affiliation{Nikhef, Science Park 105, 1098 XG Amsterdam, The Netherlands}
\author{Milan Wils\,\orcidlink{0000-0002-1544-7193}}
\affiliation{Institute for Theoretical Physics, KU Leuven, Celestijnenlaan 200D, B-3001 Leuven, Belgium}
\affiliation{Leuven Gravity Institute, KU Leuven, Celestijnenlaan 200D box 2415, 3001 Leuven, Belgium}
\author{Francesco Cireddu\,\orcidlink{0009-0002-7074-4278}}
\affiliation{Institute for Theoretical Physics, KU Leuven, Celestijnenlaan 200D, B-3001 Leuven, Belgium}
\affiliation{Dipartimento di Fisica “E. Fermi”, Università di Pisa, I-56127 Pisa, Italy}
\affiliation{INFN, Sezione di Pisa, I-56127 Pisa, Italy}
\affiliation{Leuven Gravity Institute, KU Leuven, Celestijnenlaan 200D box 2415, 3001 Leuven, Belgium}
\author{Walter Del Pozzo\,\orcidlink{0000-0003-3978-2030}}
\affiliation{Dipartimento di Fisica “E. Fermi”, Università di Pisa, I-56127 Pisa, Italy}
\affiliation{INFN, Sezione di Pisa, I-56127 Pisa, Italy}
\author{Tjonnie G. F. Li\,\orcidlink{0000-0003-4297-7365}} 
\affiliation{Institute for Theoretical Physics, KU Leuven, Celestijnenlaan 200D, B-3001 Leuven, Belgium}
\affiliation{KU Leuven, Department of Electrical Engineering (ESAT), STADIUS Center for Dynamical Systems, Signal Processing and Data Analytics, Kasteelpark Arenberg 10, 3001 Leuven, Belgium}
\affiliation{Leuven Gravity Institute, KU Leuven, Celestijnenlaan 200D box 2415, 3001 Leuven, Belgium}
\date{\today}

\begin{abstract}
    \noindent Building upon the statistical formulation for parameter estimation in the presence of correlated noise proposed by Cireddu \textit{et~al.},\
    we present the initial study to incorporate the effects of correlated noise into the analyses of various detector designs' performance.
    We consider a two-L-shaped-detector configuration located in
    Europe
    and compare the expectation of parameter estimation
    of gravitational wave transients
    between
    noncollocated and hypothetical collocated configurations.
    In our study, we posit the existence of low-frequency correlated noise within the 5--10 Hz range for the collocated detector configuration, with a varying degree of correlation. In this specific detector setup, our observations indicate an enhancement in the precision of intrinsic parameter measurements as the degree of correlation increases. This trend suggests that higher degrees of noise correlation may beneficially influence the accuracy of parameter estimation.
    In particular,
      when the noise is highly correlated,
    the uncertainty on chirp mass decreases by up to $30\%$. The absence of an inter-European baseline does hinder the estimation of the extrinsic parameters. 
    However, given a realistic global network with the additional detector located in the United States,
    the uncertainty of extrinsic parameters is significantly reduced.
    This reduction is further amplified as the degree of noise correlation increases.
    When the degree of noise correlation exceeds a certain level, the collocated configuration outperforms the noncollocated configuration.
    For instance, when the degree of correlation is high, the collocated configuration decreases the 90\% credible area of sky location by up to $10\%$ compared to the noncollocated configuration. 
    We conclude that the impact of noise correlation is not trivial and can potentially alter both the quantitative and qualitative outcomes in detector performance.
    We therefore recommend the inclusion of noise correlation for a comprehensive assessment of the design of third-generation gravitational wave detectors.
\end{abstract}

\maketitle

\section{Introduction}
\label{sec:intro}

\noindent Since the first detection of \acp{GW} in 2015~\cite{LIGOScientific:2016aoc}, almost a hundred \ac{GW} signals from binary mergers have been detected~\cite{KAGRA:2021vkt}. Such detections have significantly impacted various aspects of astronomy and fundamental physics~\cite{LIGOScientific:2021sio, KAGRA:2021duu, LIGOScientific:2023bwz}. The next-generation \ac{GW} detectors promise to unlock mysteries in astrophysics, fundamental physics, and cosmology \cite{einstein_telescope_science_case, Branchesi_2023_ET_science_case, CE_horizon_study, LISA_science_case}. One such proposal is the \ac{ET}~\cite{einstein_telescope}, a third-generation ground-based \ac{GW} detector.

One of the proposals for the \ac{ET} consists of a collocated detector network. Such a setting gives rise to a novel challenge, namely, correlated detector noise. Because of the proximity of the collocated detectors~\cite{ET_design_report}, the noise in each of the detectors is expected to have a non-negligible correlation. The correlation is likely coming from seismic perturbations, Newtonian noise \cite{janssensImpactCorrelatedSeismic2022,janssensCorrelated01Hz40HzSeismic2024}, and magnetic fluctuation \cite{janssensCorrelated11000Hz2023,janssensImpactSchumannResonances2021}.
While the literature recognizes the profound implications of correlated noise, existing research predominantly addresses its influence on the search of isotropic
stochastic
\ac{GW} background \cite{janssensImpactCorrelatedSeismic2022,janssensCorrelated11000Hz2023,janssensCorrelated01Hz40HzSeismic2024,janssensImpactSchumannResonances2021}. A recent study \cite{janssensCorrelated01Hz40HzSeismic2024} reports that the seismic correlations are significant over several hundreds of meters to a few kilometers in the frequency range 0.01--40 Hz.
Notably, the impact of correlated noise on \ac{GW} transients has not been factored into the most recent extensive evaluation of various \ac{ET} designs \cite{Branchesi_2023_ET_science_case}.
This omission highlights an important gap in our understanding and underscores the necessity for comprehensive studies that integrate the effects of correlated noise into the evaluation of detector configurations, thereby ensuring more accurate and robust design comparisons for the ET.

Conventional \ac{GW} parameter estimation \cite{veitchRobustParameterEstimation2015,ashtonBilbyUserfriendlyBayesian2019} assumes uncorrelated detector noise.
There have been studies on the \ac{LISA} that include noise correlation in their analyses \cite{1998PhRvD..57.7089C,2002PhRvD..66l2002P}.
These works acknowledge the presence of noise correlation and make specific assumptions about it.
However, the impact of noise correlation on parameter estimation was not the primary focus of these studies.
Recent work by Cireddu \textit{et~al.}\ \cite{ciredduLikelihoodNetworkGravitationalWave2023} introduces a statistical formulation to account for correlated noise
which is similar to the approach used in \ac{LISA} analyses.
This work lays a foundation for investigating the impact of noise correlation on parameter estimation, which is highly relevant for studies like ours that aim to explore these effects in more detail.

In this manuscript, we present a statistical framework to integrate the effects of correlated noise in the evaluation of different \ac{GW} detector designs.
Our approach involves numerical analyses aimed at understanding how correlated noise influences the accuracy of parameter estimation for \ac{GW} transients.
A collocated two-L-shaped configuration is often used as a proxy for a triangular configuration in the literature. However, in the
presence
of correlated noise, this is no longer generally the case.
Nevertheless, to isolate the effects of correlated noise from benefits specifically due to detector geometry, such as null stream, our study focus exclusively on two L-shaped detectors under two distinct scenarios: one in which the detectors are collocated and another where they are noncollocated.
We assess the impact on parameter estimation uncertainty through \ac{FIM} analysis to make a global assessment of a population of sources.
Remarkably, our findings reveal that, within the particular detector configuration examined, the presence of noise correlation leads to enhanced precision in parameter estimation compared to a comparable, yet noncollocated, detector network.

Hence, our research challenges the prevailing notion that noise correlation invariably compromises the scientific output of a detector network.
Instead, it furnishes compelling evidence
of the nontrivial impacts of noise correlation.
This insight not only broadens our understanding of the interplay between detector configuration and noise characteristics, but also underscores the necessity of reevaluating design strategies in light of these findings.

\section{Review of Correlated Noise Statistics}
\label{sec:review}

\noindent In this section, we review the likelihood formulation in the presence of correlated noise presented in Ref.~\cite{ciredduLikelihoodNetworkGravitationalWave2023} in the context of two detectors.

The data of the two detectors are denoted as the time series $\boldsymbol{x}_{j}$, where $j=1,2$. These time series are sampled at regular intervals of $\Delta t$. We express the time series of multiple detectors as a matrix $\boldsymbol{X}$, where each row $j$ corresponds to the time series of the $j$th detector. To compactly represent the spatial-temporal correlations in the noise data, we vectorize the matrix $\boldsymbol{X}$ into a single vector $\boldsymbol{x}$ as follows:
\begin{equation}
    \boldsymbol{x} := \textrm{vec}(\boldsymbol{X}^{T}) = 
    \begin{bmatrix}
        \boldsymbol{x}_{1} \\
        \boldsymbol{x}_{2}
    \end{bmatrix}.
\end{equation}
$\tilde{\boldsymbol{x}}$ is understood to be the vectorization of the Fourier transform of the individual time series as follows:
\begin{equation}
    \tilde{\boldsymbol{x}} = 
\begin{bmatrix}
    \tilde{\boldsymbol{x}}_{1} \\
    \tilde{\boldsymbol{x}}_{2}
\end{bmatrix}\,.
\end{equation}
The strain data are denoted as $\boldsymbol{d}$. The \ac{GW} signal is denoted as $\boldsymbol{s}(\boldsymbol{\theta})$ given the source parameters $\boldsymbol{\theta} \in \mathbb{R}^{D}$.

Assuming the noise follows the stationary Gaussian distribution,
the likelihood function can be shown to be
\begin{widetext}
\begin{equation}
\label{eq:noise_model}
    p(\boldsymbol{d} | \boldsymbol{\theta}) = 
    \frac{1}{\det(\pi\boldsymbol{S}_{n}/(2\Delta f))}
    \exp
    \left(
        -\frac{1}{2}\braket{\boldsymbol{d} - \boldsymbol{s}(\boldsymbol{\theta}),
        \boldsymbol{d} - \boldsymbol{s}(\boldsymbol{\theta})}
    \right)\,,
\end{equation}
\end{widetext}
where $\boldsymbol{S}_{n}$ is the spectral matrix defined as follows:
\begin{equation}
    \label{eq:net_psd_matrix}
    \boldsymbol{S}_{n} = 
    \begin{bmatrix}
        \boldsymbol{S}_{n}^{11} &
        \boldsymbol{S}_{n}^{12} \\
        \boldsymbol{S}_{n}^{21} &
        \boldsymbol{S}_{n}^{22} \\
    \end{bmatrix}\,,
\end{equation}
where
\begin{equation}
    \label{eq:net_psd_matrix_elements}
    \boldsymbol{S}_{n}^{\ell m}[j,k] = 2\Delta f\delta_{jk}\mathbb{E}\left[\tilde{n}_{\ell}[j]\tilde{n}_{m}^{*}[k]\right]
\end{equation}
with $\delta_{jk}$ being the Kronecker delta function. $\boldsymbol{S}_{n}^{\ell \ell}$ is the one-sided \ac{PSD} of the noise in the $\ell$th detector, and $\boldsymbol{S}_{n}^{\ell m}$ is the \ac{CSD} of the noise in the $\ell$th and the $m$th detectors. One should be reminded that our definition of \ac{CSD} differs from the conventional definition by a complex conjugate. $\braket{\boldsymbol{x},\boldsymbol{y}}$ is the generalized noise-weighed inner product defined as
\begin{equation}
    \braket{\boldsymbol{x}, \boldsymbol{y}} = 4\Delta f \textrm{Re} \left(\sum_{\ell,m=1}^{2}\sum_{k=k_{\textrm{low}}}^{k_{\textrm{high}}}(\boldsymbol{S}_{n}^{-1})^{\ell m}[k,k]\tilde{x}_{\ell}^{*}[k]\tilde{y}_{m}[k]\right),
\end{equation}
where $\Delta f$ is the frequency resolution, and the inner product evaluated between $f_{\rm low} = \Delta f k_{\rm low}$ and $f_{\rm high} = \Delta f k_{\rm high}$, where
$k_{\rm low}>0$
and $k_{\rm high} < \floor{N/2} - 1$, where $N$ is the length of the time series.

\section{Reduced Spread of the noise distribution in the presence of correlated noise}
\label{sec:reduced_spread}

\noindent While the noise model in Eq.~\eqref{eq:noise_model} accounts for the spatial-temporal correlation in the strain data of the \ac{ET}, the impact of such correlation on parameter estimation has not yet been studied. Two scenarios are considered to detail the effects of correlated noise: (i) Detectors are collocated, experiencing correlated noise, e.g., seismic noise. The corresponding noise spectral matrix is denoted as
\begin{equation}
\label{eq:cov_corr}
    \boldsymbol{S}_{n}^{\textrm{corr}} = 
    \begin{bmatrix}
        \boldsymbol{S}_{n}^{11} &
        \boldsymbol{S}_{n}^{12} \\
        \boldsymbol{S}_{n}^{21} &
        \boldsymbol{S}_{n}^{22} \\
    \end{bmatrix} \,.
\end{equation}
(ii) Detectors are noncollocated, with the extent of separation not leading to a complete alteration in environmental noise sources. Nevertheless, they still have a similar noise properties, leading to the same noise covariance matrix as scenario (i), except for the vanishing nondiagonal elements. The corresponding noise covariance matrix is given by
\begin{equation}
\label{eq:cov_uncorr}
\boldsymbol{S}_{n}^{\textrm{uncorr}} = 
    \begin{bmatrix}
        \boldsymbol{S}_{n}^{11} &
         \boldsymbol{0} \\
         \boldsymbol{0} &
        \boldsymbol{S}_{n}^{22} \\
    \end{bmatrix} \,.
\end{equation}

At this point, we can already obtain insights into the impact of correlated noise on the measurement process.
The spread of the Gaussian distribution is proportional to the square root of the determinant of the covariance or spectral matrix. From Fischer's inequality~\cite{hornMatrixAnalysis1987}, we can deduce the following inequality:
\begin{equation}
\label{eq:cov_ineq}
    \sqrt{\det{\boldsymbol{S}_{n}^{\textrm{corr}}}}
    \leq
    \sqrt{\det{\boldsymbol{S}_{n}^{\textrm{uncorr}}}},
\end{equation} 
from which one can conclude that the presence of correlated noise [i.e., nonzero off-diagonal blocks in Eq.~\eqref{eq:cov_corr}] actually reduces of the spread of the likelihood function in Eq.~\eqref{eq:noise_model}.
The propagation of the reduced spread to the precision of
parameter
estimation is, however, not trivial. We emphasize that one cannot immediately conclude from the reduced spread that it would result in a more precise parameter estimation, since a more careful analysis would reveal that the specific structure of the signal gradient is another important factor. It is, however, tempting to speculate that the impact of noise correlation is not always negative.

\vspace{0.1cm}
\section{Impact of correlated noise on parameter estimation}
\label{sec:method}
\noindent
To compare the performance,
we assess how the uncertainty of parameter estimation changes between the two detector configurations when observing the same source. This comparison is characterized by the ratio of the square root of the determinant of the covariance
matrix
of the posterior distributions, averaged over noise realizations:
\begin{widetext}
\begin{equation}
r_{\textrm{uncorr}}^{\textrm{corr}} := \frac{\mathbb{E}_{\boldsymbol{d}^{\rm corr} | \boldsymbol{\theta_{\rm true}}}\left[\sqrt{\det \left(\boldsymbol{\Sigma}_{\boldsymbol{\theta} | \boldsymbol{S}_{n}^{\textrm{corr}}}\right)}\right]}{\mathbb{E}_{\boldsymbol{d}^{\rm uncorr} | \boldsymbol{\theta_{\rm true}}}\left[\sqrt{\det \left(\boldsymbol{\Sigma}_{\boldsymbol{\theta} | \boldsymbol{S}_{n}^{\textrm{uncorr}}}\right)}\right]}= 
\frac{\int \sqrt{\det \left(\boldsymbol{\Sigma}_{\boldsymbol{\theta} | \boldsymbol{S}_{n}^{\textrm{corr}}}\right)} p(\boldsymbol{d}^{\textrm{corr}} | \boldsymbol{\theta}_{\textrm{true}}, \boldsymbol{S}_{n}^{\textrm{corr}}) \textrm{d}\boldsymbol{d}^{\textrm{corr}}}{\int \sqrt{\det \left(\boldsymbol{\Sigma}_{\boldsymbol{\theta} | \boldsymbol{S}_{n}^{\textrm{uncorr}}}\right)}p(\boldsymbol{d}^{\textrm{uncorr}} | \boldsymbol{\theta}_{\textrm{true}}, \boldsymbol{S}_{n}^{\textrm{uncorr}})  \textrm{d}\boldsymbol{d}^{\textrm{uncorr}}}
\end{equation}
\end{widetext}
where $\boldsymbol{\Sigma}_{\boldsymbol{\theta} | \boldsymbol{S}_{n}^{\textrm{corr}/\textrm{uncorr}}}$ is the posterior covariance of the model parameters $\boldsymbol{\theta}$ given a noise realization with the noise spectral matrix $\boldsymbol{S}_{n}^{\textrm{corr}/\textrm{uncorr}}$, and $\boldsymbol{\theta}_{\textrm{true}}$ is the true parameters.

Evaluating the quantity
$r_{\textrm{uncorr}}^{\textrm{corr}}$ is challenging for general setups. Instead, we turn to the \ac{FIM} analysis. The inverse of the \ac{FIM} is known to characterize the covariance of the posterior distribution in the high \ac{SNR} limit \cite{vallisneriUseAbuseFisher2008}. For a deterministic signal $\boldsymbol{s}(\boldsymbol{\theta})$, the \ac{FIM} is given by
\begin{equation}
\label{eq:fisher_matrix}
\mathcal{I}_{jk}^{\textrm{corr/uncorr}}(\boldsymbol{\theta}) = 
\left.\braket{\partial_{\theta_{j}}\boldsymbol{s}(\boldsymbol{\theta}), \partial_{\theta_{k}}\boldsymbol{s}(\boldsymbol{\theta})}\right|_{\boldsymbol{S}_{n} = \boldsymbol{S}_{n}^{\textrm{corr/uncorr}}}
\end{equation}
where $\partial_{\theta_{j}}\boldsymbol{s}(\boldsymbol{\theta})$ is the derivative of the signal with respect to the $j$th model parameter. In the high SNR limit, the ratio $r_{\textrm{uncorr}}^{\textrm{corr}}$ is then reduced to:
\begin{equation}
\label{eq:rhat}
    \hat{r}_{\textrm{uncorr}}^{\textrm{corr}} =
    \frac
    {1 / \sqrt{\det(\boldsymbol{\mathcal{I}}^{\textrm{corr}}(\boldsymbol{\theta}_{\rm true}))}}
    {1 / \sqrt{\det(\boldsymbol{\mathcal{I}}^{\textrm{uncorr}}(\boldsymbol{\theta}_{\rm true}))}}\,,
\end{equation}
$\hat{r}_{\textrm{uncorr}}^{\textrm{corr}} < 1$ is indicating an improvement in parameter estimation precision, and vice versa. The mathematical properties of $\hat{r}_{\textrm{uncorr}}^{\textrm{corr}}$ are further studied in Appendix~\ref{app:det_fisher_mat}.

\subsection{Simulation}
\noindent We consider two scenarios. In the first one, we simulate
two L-shaped
detectors
located in
Europe (EU)
with a $45^\circ$ relative offset and identical \ac{PSD}s, the ET-D sensitivity $\boldsymbol{S}_{n}^{\textrm{ET-D}}$ \cite{hildSensitivityStudiesThirdgeneration2011}. Within this scenario, we compare two configurations: (i) the two
detectors are noncollocated, with one placed at Limburg and one at Sardinia, featuring uncorrelated noise, denoted by the spectral matrix $\boldsymbol{S}_{n}^{\textrm{uncorr}}$, while for ii) the two
detectors are collocated at Limburg, featuring correlated noise, represented by the spectral matrix $\boldsymbol{S}_{n}^{\textrm{corr}}$. The orientation, elevation, and arm length are the same across the two setups. The locations and orientations of Ref.~\cite{Puecher:2023twf} are used for the
EU
detectors.

In the second scenario, we consider a simulated global network with an L-shaped detector located at Hanford, referred as the US detector in the text, with the 40 km sensitivity of the Cosmic Explorer $\boldsymbol{S}_{n}^{\textrm{CE-40km}}$ \cite{2022ApJ...931...22S,evansHorizonStudyCosmic2021,kunsCosmicExplorerStrain}.

In the simulation, we use the frequency cutoff to be $[5,1024)$ Hz. In the absence of a faithful correlated (seismic) noise model, we assume the noise components under $10$ Hz are correlated with a correlation coefficient of
$\alpha$
. Consequently, the \ac{CSD} for the EU detectors is taken to be
$\alpha S_{n}^{\textrm{ET-D}}(f)$
for $f \leq 10\textrm{ Hz}$ and $0$ otherwise.

It should be noted that the CSD,
which represent the cross covariance of noise between different detectors,
is generally complex.
This complexity arises from time delays of noise disturbances across detectors.
For simplicity,
our simulation assumes that the correlation coefficient $\alpha$ is real.
This assumption helps in focusing on the impact of varying noise correlation levels on parameter estimation.

Therefore, the spectral matrices for a sole EU network are given by
\begin{equation}
\begin{aligned}
    \boldsymbol{S}_{n}^{\textrm{uncorr}} &= 
    \begin{bmatrix}
        \boldsymbol{S}_{n}^{\textrm{ET-D}} &
        \boldsymbol{0} \\
        \boldsymbol{0} &
        \boldsymbol{S}_{n}^{\textrm{ET-D}}\\
    \end{bmatrix}, \\
    \boldsymbol{S}_{n}^{\textrm{corr}} &= 
    \begin{bmatrix}
        \boldsymbol{S}_{n}^{\textrm{ET-D}} &
        \boldsymbol{\alpha}\boldsymbol{S}_{n}^{\textrm{ET-D}} \\
        \boldsymbol{\alpha}\boldsymbol{S}_{n}^{\textrm{ET-D}} &
        \boldsymbol{S}_{n}^{\textrm{ET-D}} \\
    \end{bmatrix},
\end{aligned}
\end{equation}
and, for a global network with the inclusion of a US detector, the matrices are given by
\begin{equation}
\begin{aligned}
    \boldsymbol{S}_{n}^{\textrm{uncorr}} &= 
    \begin{bmatrix}
        \boldsymbol{S}_{n}^{\textrm{ET-D}} &
        \boldsymbol{0} &
        \boldsymbol{0} \\
        \boldsymbol{0} &
        \boldsymbol{S}_{n}^{\textrm{ET-D}} &
        \boldsymbol{0} \\
        \boldsymbol{0} &
        \boldsymbol{0} &
        \boldsymbol{S}_{n}^{\textrm{CE-40km}}
    \end{bmatrix}, \\
    \boldsymbol{S}_{n}^{\textrm{corr}} &= 
    \begin{bmatrix}
        \boldsymbol{S}_{n}^{\textrm{ET-D}} &
        \boldsymbol{\alpha}\boldsymbol{S}_{n}^{\textrm{ET-D}} &
        \boldsymbol{0} \\
        \boldsymbol{\alpha}\boldsymbol{S}_{n}^{\textrm{ET-D}}&
        \boldsymbol{S}_{n}^{\textrm{ET-D}} &
        \boldsymbol{0} \\
        \boldsymbol{0} &
        \boldsymbol{0} &
        \boldsymbol{S}_{n}^{\textrm{CE-40km}}
    \end{bmatrix},
\end{aligned}
\end{equation}
where $\alpha_{jk} = 
\alpha
\delta_{jk}H(10{\text{ Hz}} - k\Delta f)$, with $H(x)$ being the Heaviside step function.
Given the current uncertainties surrounding the level of noise correlation, we have chosen to simulate a range of correlation coefficients, $\alpha$, from $0$ $0.9$ in increments of $0.1$.
Recent measurement using seismometers at different sites and separations showed that the coherence can range between $\sim0.01$ and $\sim0.5$ at the $90$th percentile \cite{janssensCorrelated01Hz40HzSeismic2024}. The coherence cannot be trivially mapped to $\alpha$ since it depends on the exact detector layout. Nevertheless, it gives us a rough estimate of $\alpha$ in the range between $\sim0.005$ and $\sim0.25$, which is estimated by dividing the coherence by $2$ assuming only the end test mass is subject to noise correlation.

A catalog of $200$ aligned spin binary black hole merger transients using the waveform approximant IMRPhenomXPHM \cite{prattenComputationallyEfficientModels2021} is generated. The simulated catalog's source parameter distributions are outlined in Table~\ref{tab:sim_pop}. Subsequently, we perform \ac{FIM} analyses on this catalog under two different setups.

Our comparative analysis focuses on evaluating the standard deviation of individual source parameters and the spread of the posterior distributions between the two configurations under the two scenarios. It is essential to note that the \ac{FIM} only approximates the posterior primarily for high \ac{SNR} signals. Thus, we set a fiducial \ac{SNR} cutoff, ensuring signals possess an \ac{SNR} above $50$. The \ac{SNR} for this cutoff is determined by the noncollocated configuration.
\begin{table*}[htbp]
\caption{\label{tab:sim_pop}Distribution of the source parameters in the simulation. The definition of the parameters can be referred to Ref.~\cite{romero-shawBayesianInferenceCompact2020}. Cosmology from Ref.~\cite{Planck:2015fie} is used for the luminosity distance $d_{\rm L}$ prior.}
\begin{tabularx}{\textwidth}{>{\raggedright\arraybackslash}X >{\centering\arraybackslash}X >{\centering\arraybackslash}X } \toprule
    Parameter & Distribution & Range \\ \midrule
    Primary mass $m_{1}$ & Uniform & ($5$, $100$) $M_{\odot}$ \\ 
    Secondary mass $m_{2}$ & Uniform & ($5$, $100$) $M_{\odot}$ \\
    Primary aligned spin $\chi_{1}$ & Uniform & ($0$, $0.99$) \\
    Secondary aligned spin $\chi_{2}$ & Uniform & ($0$, $0.99$) \\ 
    Inclination $\theta_{\rm JN}$ & Sine & ($0$, $\pi$) \\
    Polarization angle $\psi$ & Uniform & ($0$, $\pi$) \\
    Right ascension $\alpha$ & Uniform & ($0$, $2\pi$) \\ 
    Declination $\delta$ & Cosine & ($-\pi/2$, $\pi/2$) \\ 
    Luminosity distance $d_{\rm L}$ & Uniform in source frame & ($100$, $10000$) Mpc \\
    \bottomrule
\end{tabularx}
\end{table*}

\subsection{Results}

\noindent Our findings reveal an improvement in measurement precision of the parameters
for the vast majority of the injections
when comparing
the hypothetical
collocated-correlated configuration with
the
noncollocated configuration.

\begin{table}[]
    \centering
    \caption{\label{tab:Fisher_result}Median and $1\sigma$ uncertainty range of $\hat{r}_{\textrm{uncorr}}^{\textrm{corr}}$, assessed across all, intrinsic-only, and extrinsic-only parameters comparing the collocated configuration with a correlation coefficient of $0.9$ and the noncollocated configuration.
    $\hat{r}_{\textrm{uncorr}}^{\textrm{corr}} < 1$ is indicating an improvement in parameter estimation precision, and vice versa.  Notably, the median values for intrinsic parameters consistently remain below 1, indicating increased measurement precision. In the absence of a US detector, the $\hat{r}_{\textrm{uncorr}}^{\textrm{corr}}$ values for extrinsic parameters surpass $1$. However, with a broader network with a US detector, a marked reduction below $1$ suggests superior localization precision with collocated EU detectors in a global detector network.}
    \begin{tabular}{lcc}
        \toprule
        & EU-only network & EU-US network \\ \midrule
        $\hat{r}_{\textrm{uncorr}}^{\textrm{corr}}$ (all) &
        $0.75_{-0.66}^{+19.72}$ &
        $0.20_{-0.06}^{+0.13}$ \\
        $\hat{r}_{\textrm{uncorr}}^{\textrm{corr}}$ (intrinsic) & 
        $0.38_{-0.17}^{+0.41}$ &
        $0.65_{-0.16}^{+0.18}$ \\
        $\hat{r}_{\textrm{uncorr}}^{\textrm{corr}}$ (extrinsic) & 
        $1.50_{-1.26}^{+44.17}$ &
        $0.29_{-0.07}^{+0.10}$ \\ 
        \bottomrule
    \end{tabular}
\end{table}

\begin{figure*}
    \begin{centering}
        \includegraphics[width=\linewidth]{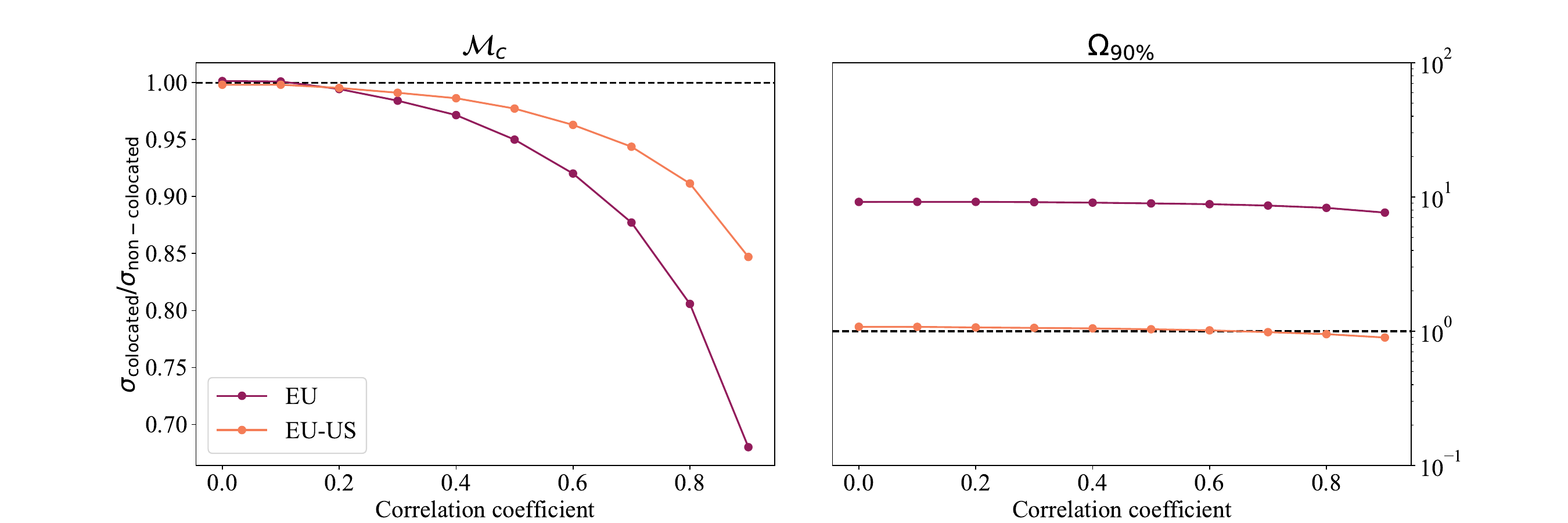}
        \caption{\label{fig:sigma_ratio_vs_corr_coef}This figure illustrates the variation in the medium of the ratio of the standard deviations of parameters between collocated and noncollocated EU detector configurations as the correlation coefficient of detector noise increases. The figure on the left displays the changes in the ratio for the chirp mass $\mathcal{M}_{c}$. In both the EU-only and EU-US detector networks, the ratio decreases with an increasing correlation coefficient, suggesting enhanced precision in parameter estimation. The right panel presents the variation in the ratio of the $90\%$ credible area for sky localization. For the EU network alone, the collocated configuration underperforms relative to the noncollocated configuration. However, with the integration of a US detector, the ratio approaches 1 and further declines to 0.9 as the noise correlation increases to 0.9, indicating improved localization accuracy with higher noise correlation.}
    \end{centering}
\end{figure*}

In Fig.~\ref{fig:sigma_ratio_vs_corr_coef},
we present a comparison of the performance variations between collocated and noncollocated European detector configurations as influenced by the correlation coefficient of detector noise. We show the results of the chirp mass as an example of the intrinsic parameters and the $90 \%$ credible area of sky localization as an example of the extrinsic parameters. In the left panel, we observe a decline in the ratio of the standard deviations for the chirp mass $\mathcal{M}_{c}$ as the correlation coefficient increases. This trend is consistent across both EU-only and EU-US detector networks, implying an improvement in the precision of parameter estimation with greater noise correlation.
The right panel presents the variation of the ratio of the $90\%$ credible area for sky localization. Here, the sole EU network exhibits poorer performance in the collocated configuration compared to the noncollocated configuration, as anticipated due to detector separation in the noncollocated case. Remarkably, the inclusion of a US detector markedly improves performance, with the ratio nearing 1 and further reducing to 0.9 as the noise correlation reaches 0.9. This showcases that the improvement from the intrinsic parameters outweighs the additional information gained from an inter-European baseline for sky localization when an intercontinental baseline is present. The distribution of the ratio for other parameters follows the same trend and is presented in Appendix~\ref{app:distribution_of_the_ratio_of_the_standard_deviations_of_all_parameters}.

To summarize the results of the \ac{FIM} analysis, Table~\ref{tab:Fisher_result} presents
the median and the $1\sigma$ interval of the ratio of posterior distributions' spreads $\hat{r}^{\rm corr}_{\rm uncorr}$ in the extreme scenario when the noise is highly correlated ($\alpha=0.9$). For both cases, the $\hat{r}_{\textrm{uncorr}}^{\textrm{corr}}$ for intrinsic parameters are consistently lower than $1$, signifying an improvement in parameter estimation precision. In the absence of
the US detector,
the $\hat{r}_{\textrm{uncorr}}^{\textrm{corr}}$ for extrinsic parameters is larger than $1$. However, with a broader network with the
US detector,
a marked reduction below $1$ suggests
improved
precision in extrinsic parameters with the collocated configuration in a global detector network.

Our results highlight the significant, yet positive, impact of noise correlation on parameter estimation accuracy within specific collocated \ac{GW} detector configurations.
This reveals the nontrivial impacts of noise correlation on the precision of parameter estimation.
This underscores the important need to consider noise correlation in the design of future GW observatories, revealing its potential to improve detection capabilities.

\section{Concluding Remarks}
\label{sec:conclusion}

\noindent Based on the statistical foundation presented in Cireddu \textit{et~al.}~\cite{ciredduLikelihoodNetworkGravitationalWave2023},
this work represents an initial endeavor to integrate the effects of correlated noise into the analysis of various detector designs' performance.
We compared the precision of parameter estimation between a hypothetical collocated two L-shaped, correlated detector
with different levels of noise correlation
with a noncollocated, uncorrelated one.
Our findings reveal a marked discrepancy in detector performance upon accounting for correlated noise impacts.
Specifically, within the considered collocated detector configuration,
rather than a hindrance, the presence of correlated noise actually boosts our ability to extract science. 

For an EU-US network setting
in the highly correlated ($\alpha\approx0.9$) scenario,
the collocated-correlated
detectors
offers a $10$\%--$20$\% improvement across parameters and an $\sim 10\%$ improvement for the sky localization, as compared to the noncollocated
detectors.

A similar study comparing different detector configurations was done in Ref.~\cite{Branchesi_2023_ET_science_case}.
Our results and Ref.~\cite{Branchesi_2023_ET_science_case}, although partially consistent, cannot be compared directly with each other because of the different configurations considered.
However, our investigations affirm that noise correlation significantly influences detector design optimization, potentially altering both quantitative and qualitative outcomes.

Our primary goal is to demonstrate that the impact of noise correlation on parameter estimation is not always negative.
It is, therefore, premature to use the zero-correlation scenario as the upper bound for performance in correlated scenarios to simplify the analysis.

In conclusion, despite its importance, the influence of correlated noise remains overlooked in current discussions on future \ac{GW} detector design. We recommend
a thorough consideration of noise correlation in future research endeavors, emphasizing its potential to reshape our understanding and approach to detector design optimization in the pursuit of advancing GW astronomy.

\section*{Acknowledgement}

\noindent We thank Rico K.~L.\ Lo for insightful discussions. 
This work was partially supported by the Research Foundation - Flanders (Grant No. I002123N).
P.~T.~H.~P is supported by the research program of the Netherlands Organization for Scientific Research (NWO).
M.W. is supported by the Research Foundation - Flanders
(FWO) through Grant No.~11POK24N.

\section*{Software}
\noindent The analysis and simulations presented in this work were conducted using a combination of established software packages and custom scripts. The figures are produced using {\sc Matplotlib} \cite{hunterMatplotlib2DGraphics2007,the_matplotlib_development_team_2023_10150955} and {\sc seaborn} \cite{waskomSeabornStatisticalData2021}. Numerical operations and array manipulations are performed using {\sc NumPy} \cite{harrisArrayProgrammingNumPy2020} and {\sc SciPy} \cite{virtanenSciPyFundamentalAlgorithms2020}. To evaluate the Fisher information matrix, {\sc gwbench}\cite{borhanianGwbenchNovelFisher2021} is used with extension to analyze scenarios involving correlated noise.

\section*{Data availability}
\noindent The data generated for this publication is available on Zenodo at Ref.~\cite{wong_2025_14842436}.
The postprocessing scripts used to analyze and visualize the results are archived with Zenodo at Ref.~\cite{isaac_c_f_wong_2025_14844558}.
The scripts used to generate the data are not publicly available, but we are happy to share them upon request.
Please note that they may require additional cleaning and documentation.

\begin{acronym}
    \acro{GW}[GW]{gravitational wave}
    \acro{ET}[ET]{Einstein Telescope}
    \acro{SNR}[SNR]{signal-to-noise ratio}
    \acro{LF}[LF]{low-frequency}
    \acro{HF}[HF]{high-frequency}
    \acro{ASD}[ASD]{amplitude spectral density}
    \acro{PSD}[PSD]{power spectral density}
    \acro{CSD}[CSD]{cross spectral density}
    \acro{BBH}[BBH]{binary black hole}
    \acro{FIM}[FIM]{Fisher information matrix}
    \acro{PE}[PE]{parameter estimation}
    \acro{CE}[CE]{Cosmic Explorer}
    \acro{LISA}[LISA]{Laser Interferometer Space Antenna}
\end{acronym}

\appendix

\section{Determinant of the Inverse Fisher Information Matrix}
\label{app:det_fisher_mat}

\noindent The \ac{FIM} is defined as
\begin{equation}
    \mathcal{I}_{jk}(\boldsymbol{\theta}) = 
    \mathbb{E}_{\boldsymbol{d} | \boldsymbol{\theta}}
    \left[
        (\partial_{\theta_{j}}\log p(\boldsymbol{d} | \boldsymbol{\theta}))
        (\partial_{\theta_{k}}\log p(\boldsymbol{d} | \boldsymbol{\theta}))
    \right] \,.
\end{equation}
With the time domain likelihood function defined in Ref.~\cite{ciredduLikelihoodNetworkGravitationalWave2023}, one can show that
\begin{equation}
    \boldsymbol{\mathcal{I}}(\boldsymbol{\theta}) = (\nabla \boldsymbol{s}(\boldsymbol{\theta}))^{T}\boldsymbol{\Sigma}_{n}^{-1}\nabla\boldsymbol{s}(\boldsymbol{\theta})
\end{equation}
where $\nabla \boldsymbol{s}(\boldsymbol{\theta})$ is the Jacobian matrix of $\boldsymbol{s}(\boldsymbol{\theta})$ and $\boldsymbol{\Sigma}_{n}$ is the noise covariance matrix.
Perform a singular value decomposition on $\nabla\boldsymbol{s}(\boldsymbol{\theta})$, we have
\begin{equation}
    \nabla\boldsymbol{s}(\boldsymbol{\theta}) = 
    \boldsymbol{U}(\boldsymbol{\theta})\boldsymbol{S}(\boldsymbol{\theta})(\boldsymbol{V}(\boldsymbol{\theta}))^{T}
\end{equation}
where $\boldsymbol{U}(\boldsymbol{\theta}) \in \mathbb{R}^{MN\times MN}$ and $\boldsymbol{V}(\boldsymbol{\theta})\in \mathbb{R}^{D\times D}$ are orthogonal  matrices, and $\boldsymbol{S}(\boldsymbol{\theta}) \in \mathbb{R}^{MN\times D}$ is a rectangular matrix with the singular values in the principal diagonal. $M$ is the number of detectors, $N$ is the number of time bins, and $D$ is the dimensionality of the parameter space $\boldsymbol{\Theta}$. Substitute the decomposition into $\hat{r}_{\textrm{uncorr}}^{\textrm{corr}}$ defined in Eq.~\eqref{eq:rhat}, we have
\begin{widetext}
\begin{align}
    &(\hat{r}_{\textrm{uncorr}}^{\textrm{corr}})^{2} \\
    &=
    \left.
    \frac
    {
    \det
    \left(
    (\nabla_{\boldsymbol{\theta}}
    \boldsymbol{s}(\boldsymbol{\theta})
    )^{T}
    (\boldsymbol{\Sigma}_{n}^{\textrm{uncorr}})^{-1}
    (\nabla_{\boldsymbol{\theta}}
    \boldsymbol{s}(\boldsymbol{\theta})
    )
    \right)
    }
    {
    \det
    \left(
    (\nabla_{\boldsymbol{\theta}}
    \boldsymbol{s}(\boldsymbol{\theta})
    )^{T}
    (\boldsymbol{\Sigma}_{n}^{\textrm{corr}})^{-1}
    (\nabla_{\boldsymbol{\theta}}
    \boldsymbol{s}(\boldsymbol{\theta})
    )
    \right)
    }
    \right|_{\boldsymbol{\theta} = \boldsymbol{\theta}_\textrm{true}} \\
    &=
    \left.\frac
    {\det
    \left(
    \boldsymbol{V}(\boldsymbol{\theta})
    \boldsymbol{S}(\boldsymbol{\theta})^{T}
    \boldsymbol{U}(\boldsymbol{\theta})^{T}
    (\boldsymbol{\Sigma}_{n}^{\textrm{uncorr}})^{-1}
    \boldsymbol{U}(\boldsymbol{\theta})
    \boldsymbol{S}(\boldsymbol{\theta})
    \boldsymbol{V}(\boldsymbol{\theta})^{T}
    \right)
    }
    {
    \det
    \left(
    \boldsymbol{V}(\boldsymbol{\theta})
    \boldsymbol{S}(\boldsymbol{\theta})^{T}
    \boldsymbol{U}(\boldsymbol{\theta})^{T}
    (\boldsymbol{\Sigma}_{n}^{\textrm{corr}})^{-1}
    \boldsymbol{U}(\boldsymbol{\theta})
    \boldsymbol{S}(\boldsymbol{\theta})
    \boldsymbol{V}(\boldsymbol{\theta})^{T}
    \right)
    }\right|_{\boldsymbol{\theta}=\boldsymbol{\theta}_{\textrm{true}}} \\
    \label{eq:rhat_deri_1}
    &=
    \left.\frac
    {
    \left(
    \det \boldsymbol{V}(\boldsymbol{\theta})
    \right)^{2}
    \det
    \left(
    \boldsymbol{S}(\boldsymbol{\theta})^{T}
    \boldsymbol{U}(\boldsymbol{\theta})^{T}
    (\boldsymbol{\Sigma}_{n}^{\textrm{uncorr}})^{-1}
    \boldsymbol{U}(\boldsymbol{\theta})
    \boldsymbol{S}(\boldsymbol{\theta})
    \right)
    }
    {
    \left(
    \det \boldsymbol{V}(\boldsymbol{\theta})
    \right)^{2}
    \det
    \left(
    \boldsymbol{S}(\boldsymbol{\theta})^{T}
    \boldsymbol{U}(\boldsymbol{\theta})^{T}
    (\boldsymbol{\Sigma}_{n}^{\textrm{corr}})^{-1}
    \boldsymbol{U}(\boldsymbol{\theta})
    \boldsymbol{S}(\boldsymbol{\theta})
    \right)
    }\right|_{\boldsymbol{\theta}=\boldsymbol{\theta}_{\textrm{true}}} \\
    \label{eq:rhat_deri_2}
    &=
    \left.\frac
    {
    \det
    \left(
    \hat{\boldsymbol{S}}(\boldsymbol{\theta})^{T}
    \left[
    \boldsymbol{U}(\boldsymbol{\theta})^{T}(\boldsymbol{\Sigma}_{n}^{\textrm{uncorr}})^{-1}
    \boldsymbol{U}(\boldsymbol{\theta})
    \right]_{\alpha,\alpha}
    \hat{\boldsymbol{S}}(\boldsymbol{\theta})
    \right)
    }
    {
    \det
    \left(
    \hat{\boldsymbol{S}}(\boldsymbol{\theta})^{T}
    \left[
    \boldsymbol{U}(\boldsymbol{\theta})^{T}(\boldsymbol{\Sigma}_{n}^{\textrm{corr}})^{-1}
    \boldsymbol{U}(\boldsymbol{\theta})
    \right]_{\alpha,\alpha}
    \hat{\boldsymbol{S}}(\boldsymbol{\theta})
    \right)
    }\right|_{\boldsymbol{\theta}=\boldsymbol{\theta}_{\textrm{true}}} \\
    &=
    \left.\frac
    {
    \left(
    \det\hat{\boldsymbol{S}}(\boldsymbol{\theta})
    \right)^{2}
    \det
    \left(
    \left[
    \boldsymbol{U}(\boldsymbol{\theta})^{T}(\boldsymbol{\Sigma}_{n}^{\textrm{uncorr}})^{-1}
    \boldsymbol{U}(\boldsymbol{\theta})
    \right]_{\alpha,\alpha}
    \right)
    }
    {
    \left(
    \det\hat{\boldsymbol{S}}(\boldsymbol{\theta})
    \right)^{2}
    \det
    \left(
    \left[
    \boldsymbol{U}(\boldsymbol{\theta})^{T}(\boldsymbol{\Sigma}_{n}^{\textrm{corr}})^{-1}
    \boldsymbol{U}(\boldsymbol{\theta})
    \right]_{\alpha,\alpha}
    \right)
    }\right|_{\boldsymbol{\theta}=\boldsymbol{\theta}_{\textrm{true}}} \\
    &=
    \left.\frac
    {\det
    \left(
    \left[
    \boldsymbol{U}(\boldsymbol{\theta})^{T}(\boldsymbol{\Sigma}_{n}^{\textrm{uncorr}})^{-1}
    \boldsymbol{U}(\boldsymbol{\theta})
    \right]_{\alpha,\alpha}
    \right)}
    {\det
    \left(
    \left[
    \boldsymbol{U}(\boldsymbol{\theta})^{T}(\boldsymbol{\Sigma}_{n}^{\textrm{corr}})^{-1}
    \boldsymbol{U}(\boldsymbol{\theta})
    \right]_{\alpha,\alpha}
    \right)}\right|_{\boldsymbol{\theta}=\boldsymbol{\theta}_{\textrm{true}}} \\
    &=
    \left.\frac
    {1 / \det
    \left(
    \left[
    \boldsymbol{U}(\boldsymbol{\theta})^{T}(\boldsymbol{\Sigma}_{n}^{\textrm{corr}})^{-1}
    \boldsymbol{U}(\boldsymbol{\theta})
    \right]_{\alpha,\alpha}
    \right)}
    {
    1 / \det
    \left(
    \left[
    \boldsymbol{U}(\boldsymbol{\theta})^{T}(\boldsymbol{\Sigma}_{n}^{\textrm{uncorr}})^{-1}
    \boldsymbol{U}(\boldsymbol{\theta})
    \right]_{\alpha,\alpha}
    \right)}\right|_{\boldsymbol{\theta}=\boldsymbol{\theta}_{\textrm{true}}}
    \label{eq:rhat_U}
\end{align}
\end{widetext}
where $\alpha = \{1, 2, ..., D\}$ denotes the index set, and $\left[ {}\cdot{} \right]_{\alpha,\alpha}$ represents the submatrix with the index set $\alpha$. It is noteworthy that the spread of the posterior covariance matrix is inversely proportional to $\sqrt{\det \left( \left[ \boldsymbol{U}(\boldsymbol{\theta})^{T}(\boldsymbol{\Sigma}_{n}^{\textrm{corr/uncorr}})^{-1} \boldsymbol{U}(\boldsymbol{\theta}) \right]_{\alpha,\alpha} \right)}$. From Eq.~\eqref{eq:rhat_deri_1} to Eq.~\eqref{eq:rhat_deri_2}, we replace $\boldsymbol{S}(\boldsymbol{\theta})$ with $\hat{\boldsymbol{S}}(\boldsymbol{\theta})$ defined as a square matrix with the singular values as the diagonal entries.

The geometrical interpretation is clear: the uncertainty in parameter estimation is inversely proportional to the spread of the subspace spanned by the $D$ principal vectors in the inverse noise covariance matrix. These vectors represent the directions in which the signal varies the most. A larger subspace spread corresponds to reduced uncertainty in estimating the model parameters.

In Eq.~\eqref{eq:rhat_U}, one shall notice that the singular values are
canceled,
implying that the ratio is independent of the strength of the signal. Moreover, the uncertainty is determined by the subspace spanned by the signal gradient vectors. Provided that the subspaces $\boldsymbol{U}$ are the same in the numerator and the denominator, the ratio of the uncertainty is largely dependent on the spread of the noise covariance matrix. Thus, Eq.~\eqref{eq:rhat_U} suggests that the reduced spread of the noise distribution is likely to lead to a smaller uncertainty in parameter estimation of \ac{GW} transients.

\begin{figure*}
    \begin{centering}
        \includegraphics[width=\linewidth]{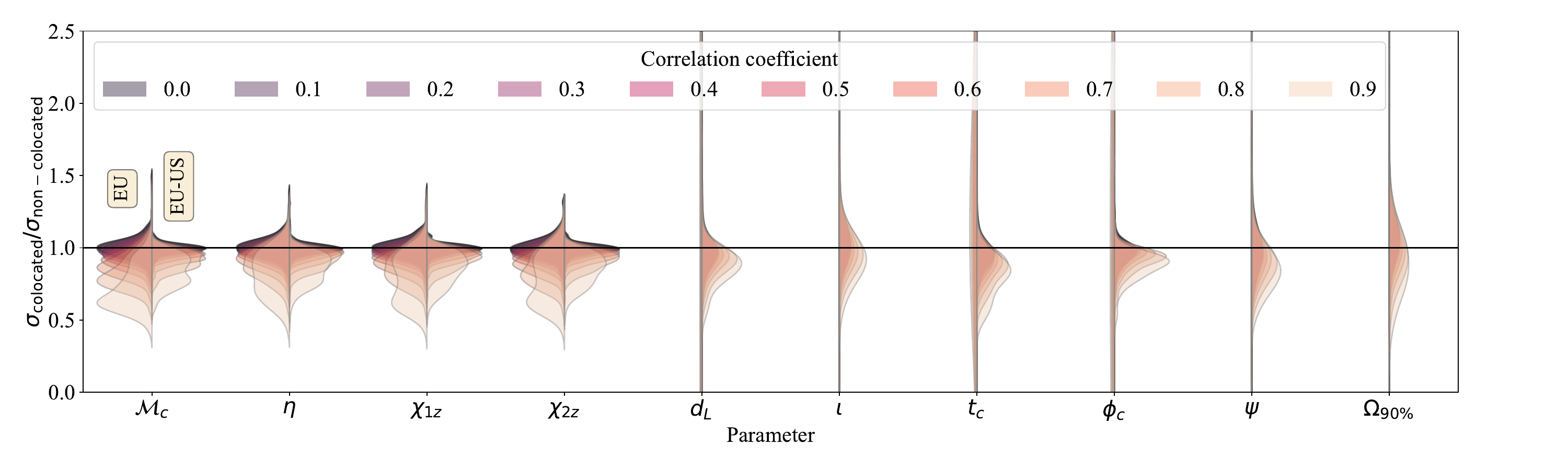}
        \caption{The distribution of the ratio of the standard deviations of the individual parameters of the posterior distributions,
        with different correlation coefficients,
        including the chirp mass $\mathcal{M}_{c}$, symmetric mass ratio $\eta$, aligned spins $\chi_{1,2z}$, coalescence phase $\phi_{c}$, coalescence time $t_{c}$, polarization angle $\psi$, inclination angle $\iota$, and luminosity distance $d_{L}$, comparing the colocated configuration and the non-colocated configuration
        of the EU detectors.
        LHS of the plots presents the results without including the US detector.
        RHS of the plots presents the results including the US detector.
        The ratio of the $90\%$ credible area of the sky localization $\Omega_{90\%}$ is presented in the last column.}
        \label{fig:fisher_csize}
    \end{centering}
\end{figure*}

\section{Distribution of the ratio of the standard deviations of all parameters}
\label{app:distribution_of_the_ratio_of_the_standard_deviations_of_all_parameters}

In Figure~\ref{fig:fisher_csize}, we present the distributions of the ratio of the standard deviations of the individual parameters comparing the colocated configuration with different correlation coefficients and the non-colocated configuration. The left-hand side of the violin plots showcases outcomes from Europe-only network, while the right-hand side illustrates findings from an Europe-US network.

In the Europe-only network,
as the correlation coefficient increases, the measurement uncertainty of the intrinsic parameters decreases.
In particular,
when the noise is highly correlated ($\alpha=0.9$),
the uncertainty of chirp mass $\mathcal{M}_c$ decreases by $\sim 30\%$, with a $\sim 10\%$ decrease for the symmetric mass ratio $\eta$ and the component aligned spins $\chi_{1z, 2z}$. For the extrinsic parameters, as expected,
the non-colocated
network
is offering better estimation.
Yet, it would be more realistic to consider a global detector network.

With the inclusion of the US detector at Hanford, as shown on the right-hand side of the plots, the colocated configuration improves the measurement precision of all parameters compared to the non-colocated configuration. The degree of improvement increases with the correlation coefficient.
Notably, when the noise is highly correlated ($\alpha = 0.9$), the uncertainties for most of the parameters decrease by $\sim 10 - 20\%$, and the area of the $90\%$ credible interval of the sky location $\Omega_{90\%}$ decreases by $\sim 50\%$. In Table~\ref{tab:parameter_sigma_without_CE} and Table~\ref{tab:parameter_sigma_with_CE}, the median and $1\sigma$ intervals of the same distributions in the highly correlated scenario ($\alpha=0.9$) are shown.

\begin{table*}
\caption{\label{tab:parameter_sigma_without_CE}The median and the 1$\sigma$ interval of the ratio of the standard deviations of the individual parameters in different detector networks compared to the non-colocated configuration, without the presence of the US detector. The colocated-uncorrelated EU network acts as a control to compare with colocated-correlated EU network with a correlation coefficient $\alpha$ of $0.9$ to examine the impact of the presence of correlated noise.}
\begin{tabular}{lcccccccccc}
\toprule
\multicolumn{1}{l}{\textbf{EU-only}} & \multicolumn{10}{c}{} \\ \midrule
Detector network & $\mathcal{M}_{c}$ & $\eta$ & $\chi_{1z}$ & $\chi_{2z}$ & $\phi_{c}$ & $t_{c}$ & $\psi$ & $\iota$ & $d_{L}$ & $\Omega_{90\%}$   \\ \midrule
colocated-correlated  &
$0.68_{-0.10}^{+0.22}$ &
$0.82_{-0.19}^{+0.15}$ & 
$0.77_{-0.19}^{+0.19}$ & 
$0.77_{-0.18}^{+0.20}$ & 
$1.22_{-0.41}^{+1.52}$ & 
$1.41_{-0.62}^{+8.73}$ & 
$1.63_{-0.75}^{+5.58}$ & 
$0.92_{-0.12}^{+0.34}$ & 
$1.16_{-0.37}^{+3.54}$ & 
$7.65_{-6.68}^{+337.72}$
\\
colocated-uncorrelated &
$1.00_{-0.06}^{+0.05}$ &
$1.00_{-0.06}^{+0.06}$ & 
$1.01_{-0.07}^{+0.05}$ & 
$1.00_{-0.06}^{+0.07}$ & 
$1.37_{-0.41}^{+1.73}$ & 
$1.55_{-0.64}^{+10.06}$ & 
$1.81_{-0.81}^{+6.54}$ & 
$1.04_{-0.09}^{+0.37}$ & 
$1.34_{-0.43}^{+4.30}$ & 
$9.19_{-7.91}^{+378.12}$
\\
non-colocated & 1 & 1 & 1 & 1 & 1 & 1 & 1 & 1 & 1 & 1 \\ \bottomrule
\end{tabular}
\end{table*}

\begin{table*}
\caption{\label{tab:parameter_sigma_with_CE}The median and the 1$\sigma$ interval of the ratio of the standard deviations of the individual parameters in different detector networks compared to the non-colocated configuration, with the presence of the US detector. The colocated-uncorrelated EU network acts as a control to compare with colocated-correlated EU network with a correlation coefficient $\alpha$ of $0.9$ to examine the impact of the presence of correlated noise.}
\begin{tabular}{lcccccccccc}
\toprule
\multicolumn{1}{l}{\textbf{EU-US}} & \multicolumn{10}{c}{} \\ \midrule
Detector network & $\mathcal{M}_{c}$ & $\eta$ & $\chi_{1z}$ & $\chi_{2z}$ & $\phi_{c}$ & $t_{c}$ & $\psi$ & $\iota$ & $d_{L}$ & $\Omega_{90\%}$   \\ \midrule
colocated-correlated &
$0.85_{-0.11}^{+0.11}$ &
$0.87_{-0.13}^{+0.10}$ & 
$0.86_{-0.14}^{+0.10}$ & 
$0.86_{-0.13}^{+0.11}$ & 
$0.82_{-0.20}^{+0.11}$ & 
$0.93_{-0.13}^{+0.13}$ & 
$0.85_{-0.14}^{+0.13}$ & 
$0.89_{-0.12}^{+0.07}$ & 
$0.89_{-0.13}^{+0.07}$ & 
$0.90_{-0.21}^{+0.23}$
\\
colocated-uncorrelated &
$1.00_{-0.03}^{+0.03}$ &
$0.99_{-0.02}^{+0.03}$ & 
$1.00_{-0.03}^{+0.03}$ & 
$1.00_{-0.03}^{+0.03}$ & 
$0.95_{-0.08}^{+0.08}$ & 
$1.03_{-0.06}^{+0.10}$ & 
$0.99_{-0.12}^{+0.14}$ & 
$0.98_{-0.08}^{+0.05}$ & 
$0.98_{-0.09}^{0.06}$ & 
$1.08_{-0.10}^{+0.13}$
\\
non-colocated & 1 & 1 & 1 & 1 & 1 & 1 & 1 & 1 & 1 & 1 \\ \bottomrule
\end{tabular}
\end{table*}

\clearpage
\bibliography{bib}

\end{document}